\documentclass{elsart}
\usepackage{fleqn}
\usepackage{graphicx}
\usepackage{epsfig}
\usepackage{amsmath}
\usepackage{amsfonts}
\usepackage{amssymb}

\def\be{\begin{equation}}
\def\ee{\end{equation}}
\def\bea{\begin{eqnarray}}
\def\eea{\end{eqnarray}}
\def\ep{\epsilon}

\def\T{{\mathcal F}}
\def\N{{\mathcal G}}
\def\TT{{\mathcal T}}
\def\NN{{\mathcal V}}
\def\t{\tau_z}

\begin{document}

\begin{frontmatter}

\title{\bf Simple solutions of relativistic hydrodynamics\\
	for cylindrically symmetric systems}

\author[KFKI,USP]{T. Cs\"org\H o,\thanksref{tamas}}
\author[USP]{F. Grassi,\thanksref{frederique}}
\author[USP]{Y. Hama,\thanksref{yogiro}}
\author[UFRJ]{T.  Kodama\thanksref{takeshi}}
\address[KFKI]{MTA KFKI RMKI, H - 1525 Budapest 114, POBox 49, Hungary}
\address[USP]{IF-USP, C.P. 66318, 05389-970 S\~{a}o Paulo, SP, Brazil}
\address[UFRJ]{IF-UFRJ, C.P. 68528, 21945-970 Rio de Janeiro, RJ, Brazil}
\thanks[tamas]{Email: csorgo@sunserv.kfki.hu}
\thanks[frederique]{\phantom{Email:} grassi@if.usp.br}
\thanks[yogiro]{\phantom{Email:} hama@fma.if.usp.br}
\thanks[takeshi]{\phantom{Email:} tkodama@if.ufrj.br}

\date{Febr. 10. 2002}

\begin{abstract}
Simple, self-similar, analytic solutions of 1 + 3 dimensional
relativistic hydrodynamics are presented for cylindrically 
symmetric fireballs corresponding to
central collisions of heavy ions at relativistic bombarding
energies.
\end{abstract}

\end{frontmatter}

\date{28.07.2001}
\maketitle

\section{Introduction}
The analytic resolution of 3 dimensional relativistic hydrodynamics 
is a difficult task because the equations are non-linear and the
transverse and the longitudinal motions of the fluid are coupled 
in a way that is rather complicated to handle. 

Historically, the first solutions of relativistic hydrodynamics were 
found for idealized, one dimensionally expanding sources: Landau's 
one-dimensional analytical solution of relativistic hydrodynamics 
\cite{Landau} and Bjorken's boost-invariant solution \cite{Bjorken}, 
that correspond to ultrarelativistic bombarding energies. Both of 
these solutions are frequently utilized  for various estimations of 
observables in ultra-relativistic nucleus-nucleus collisions. 

The repertoir of relativistic hydrodynamics has been reviewed 
from the point of view of heavy ion physics in ref.~\cite{csernai}. 
Recently, Bir\'o has found  self-similar exact solutions of 
relativistic hydrodynamics for cylindrically expanding systems
\cite{biro1,biro2}.
Bjorken's one-dimensional solution can be naturally
extended to a three-dimensional solution, assuming spherical symmetry instead of (1+1) dimensional longitudinal dynamics. 
Bir\'o's solutions generalize the (1+3) dimensional version of 
Bjorken's solution to the case of cylindrical symmetry, 
interpolating between the one-dimensional and the three-dimensional 
Bjorken solution. However, Bir\'o's solutions are valid only when 
the pressure is independent of space and time, as happens at the 
softest point of the equation of state. 

The numerical solution of the 1+3 dimensional relativistic 
hydrodynamics is possible with presently available computers and 
methods, for example the variational method of smoothed particle 
hydrodynamics has been developed to study event-by-event 
fluctuations in relativistic heavy-ion 
collisions~\cite{SPH-jpg,SPH-qm01}. 

Here we describe a new family of exact analytic solutions of 
relativistic hydrodynamics with a broad family of equations of 
state, assuming cylindrical symmetry of the particle emitting 
source. The physical motivation for this study is to consider the 
time evolution of central collisions in ultra-relativistic 
heavy-ion physics within the framework of an analytic approach. 

Our solution was motivated by the analytic solution of 
non-relativistic hydrodynamics found by Zim\'anyi, Bondorf and 
Garpman (ZBG) in 1978 for low-energy heavy-ion collisions with 
spherical symmetry~\cite{jnr}. This solution has been generalized 
to the case of ellipsoidal symmetry by Zim\'anyi and collaborators 
in ref.~\cite{jde}. In ref. ~\cite{nr,nrt} a Gaussian 
parameterization has been introduced to describe the mass dependence 
of the effective temperature and the radius parameters of the 
two-particle Bose-Einstein correlation functions in high-energy 
heavy-ion collisions. Later it has been realized that this 
phenomenological {\it parameterization} of data corresponds to an 
exact, Gaussian {\it  solution} of non-relativistic hydrodynamics 
with spherical symmetry~\cite{cspeter}. The spherically symmetric 
self-similar solutions of non-relativistic hydrodynamics were 
generalized in ref.~\cite{cssol} in terms of an arbitrary scaling 
function for the temperature profile, and expressing the density 
distribution in terms of this scaling function. The spherically 
symmetric Gaussian solution has been generalized to ellipsoidally 
symmetric expansions~\cite{ellsol}, that provided simple analytic 
insight into the observables of non-central heavy-ion 
collisions~\cite{ellsp}. The family of ellipsoidally symmetric 
solutions of non-relativistic hydrodynamics has been expressed in 
terms of a general scaling function for the temperature profile in 
ref.~\cite{csell}. 

Our approach corresponds to a generalization of these recently 
obtained analytic solutions~\cite{cspeter,cssol,ellsp,csell} 
of non-relativistic fireball hydrodynamics to the case of 
relativistic longitudinal and transverse flows. In particular, an 
analytic approach, the Buda-Lund (BL) model has been developed to 
{\it parameterize} the single particle spectra and the two-particle 
Bose-Einstein correlations in high-energy heavy-ion physics in 
terms of hydrodynamically expanding, cylindrically symmetric 
sources~\cite{3d}. Here we attempt to find an exact {\it family of 
solutions} of relativistic hydrodynamics that may include the BL 
model as a particular limiting case. 

We attempt here to solve the full (1+3) dimensional relativistic 
hydrodynamical problem, in trying to overcome two shortcomings of 
Bjorken's well known solution. These two shortcomings of Bjorken's 
solutions are that {\it i)} it describes ultra-relativistic 
collisions, it is independent of scale parameter in the longitudinal 
direction and the rapidity distribution is flat; {\it ii)} it 
contains no transverse flow.
 
We have described recently a new family of (1+1) dimensional 
solutions~\cite{s1} that is able to describe an arbitrary rapidity 
distribution $dn/dy$ by introducing scale-dependent quantities in 
the longitudinal direction, overcoming shortcoming ${\it i)}$, but 
not addressing shortcoming ${\it ii)}$. Here we consider the 
opposite case, overcoming shortcoming {\it ii)}, the development of 
a scale dependent transverse flow in relativistic hydrodynamics, 
but not addressing shortcoming {\it i)}. We address both 
shortcomings simultaneously in a subsequent 
manuscript~\cite{rellsol}.
 
\section{The equations of relativistic hydrodynamics} 
We solve the relativistic continuity and energy-momentum conservation
equation: 
\begin{eqnarray}
\partial_\mu(n u^\mu) & = & 0\,, \label{e:cont} \\
\partial_\nu T^{\mu \nu} & = & 0\,.  \label{e:tnm} 
\end{eqnarray}
Here the $n \equiv n(t,{\mathbf r})$ is the number density, the 
four-velocity is denoted by 
$u^\mu \equiv u^\mu(t,{\mathbf r}) = \gamma (1, {\mathbf v})$, 
normalized to $u^\mu u_\mu = \gamma^2 (1 - {\mathbf v}^2) = 1$, and 
the energy-momentum tensor is denoted by $T^{\mu \nu}$. We assume 
perfect fluid, 
\begin{equation} 
T^{\mu\nu} = (\ep + p) u^\mu u^\nu - p g^{\mu \nu},
\end{equation} 
where $\ep$ stands for the relativistic energy density and $p$ denotes the pressure.

We close this set of relativistic hydrodynamical equations 
with the equations of state. We assume a gas containing massive 
conserved quanta, 
\bea
\ep & =& m n + \kappa p, \label{e:eos1}\\
p  & = & n T \label{e:eos2}.
\eea
The equations of state have two free parameters, $m$ and $\kappa$.
Non-relativistic hydrodynamics of ideal gases corresponds to the
limiting case  of 
$m \gg T$, ${\mathbf v}^2 \ll 1$ and $\kappa = \frac{3}{2}$.

The energy-momentum conservation equations can be projected into a 
component parallel to $u^\mu$ and components orthogonal to $u^\mu$, 
which are respectively the relativistic energy and Euler equations: 
\bea
u^\mu \partial_\mu \ep + (\ep + p) \partial_\mu u^\mu & = & 0\,, \label{e:ren}\\
u_\nu u^\mu \partial_\mu p + (\ep + p)u^\mu \partial_\mu u_\nu - 
\partial_\nu p & = & 0\,. \label{e:rEu}
\eea 

With the help of the equations of state and the continuity equation, 
the energy equation  can be rewritten as an equation for the 
temperature, 
\begin{equation} 
u^\mu \partial_\mu T + \frac{1}{\kappa} T \partial_\mu u^\mu = 0\,. \label{e:rT}
\end{equation} 
We adopt the following notational conventions:
the coordinates are $x^\mu = (t, {\mathbf r}) = (t, r_x, r_y, r_z)$,
$x_\mu = (t, -r_x, -r_y, -r_z)$ and the metric tensor is 
$g^{\mu\nu} = g_{\mu\nu} = \mbox{\rm diag}(1,-1,-1,-1)$. 

We solve 5 independent equations, the continuity, 
the (3 spatial  components of) relativistic Euler, 
and the temperature equation, 
eqs.~(\ref{e:cont},\ref{e:rEu},\ref{e:rT}).
The equations of state, eq.~(\ref{e:eos1},\ref{e:eos2}) close this system of equations in terms of 5 variables, $n$, $T$ and 
${\mathbf v} = (v_x,v_y,v_z)$. 
 
\section{\it Self-similar, cylindrically symmetric solutions}
As we are primarily interested in the effects of finite transverse 
size and the development of transverse flow, we assume that the 
longitudinal flow component is that of Bjorken's type, 
\begin{equation} 
v_z(t,r_z)  =  \frac{r_z}{t} \label{e:vz}.
\end{equation} 
A similar assumption has been made when a new family of (1+1) 
dimensional  solutions of relativistic hydrodynamics was obtained 
in ref.~\cite{s1}. However, in contrast to what has been done in 
\cite{s1}, here we assume scale-invariance in the longitudinal 
direction. 

We search for self-similar solutions, that are scale dependent in 
the transverse directions, and depend only on the transverse radius 
variable $r_t = \sqrt{r_x^2 + r_y^2}$ and the longitudinal proper 
time $\t = \sqrt{t^2 - r_z^2}\,$. Let us introduce the scaling 
variable $x$ as 
\begin{equation} 
	x=\frac{r_x^2 + r_y^2}{R^2}, 
\end{equation} 
and assume that, in the frame where $v_z=0$ (longitudinal proper frame), the transverse motion corresponds to a Hubble type of 
self-similar transverse expansion, 
\bea
v_x^*(\t,r_z)& = & {\frac{\dot{R}(\t)}{R(\t)}}r_x\,, \label{e:vx}\\ 
v_y^*(\t,r_z) & = & {\frac{\dot{R}(\t)}{R(\t)}}r_y\,, \label{e:vy} 
\eea
where $\dot R = dR(\t)/d\t$ and hereafter we will designate by 
starred symbols the variables in the longitudinal proper frame. We 
assume that the scale $R$ depends on time only trough the 
longitudinal proper time, $\t$. 

In a relativistic notation, the above form may be parametrized as  
\bea
u^\mu & = & (\cosh\zeta \cosh\xi, 
		\sinh\xi \frac{r_x}{r_t}, \sinh\xi \frac{r_y}{r_t},
		\sinh\zeta \cosh\xi),\\ 
\tanh\xi& = &{\frac{\dot{R}(\t)}{R(\t)}}r_t=v_t^*=\gamma_lv_t\,, \\ 
\cosh\xi& = &\frac{1}{\sqrt{1 - \dot R^2 x}} \equiv \gamma_t^*\,,\\ 
\cosh\zeta & = & \frac{t}{\t} \equiv \gamma_l\,.
\eea
The space-time rapidity $\eta$ is defined as
\begin{equation} 
\eta  =  0.5 \log\left(\frac{t + r_z}{t - r_z}\right). \label{e:eta}
\end{equation} 
For a scaling longitudinal Bjorken flow we obtain
\begin{equation} 
\zeta = \eta\,. 
\end{equation}  
Using the above ansatz for the flow velocity distribution, 
we find that the continuity equation is solved by the form
\begin{equation} 
n(t,r_x,r_y,r_z) = n_0 
	\left(\frac{\tau_{z0}R_0^2}{\t R^2}\right)
	\frac{1}{\cosh\xi} \N(x), \label{s:cont}
\end{equation} 
where $\N(x)$ is an arbitrary non-negative function of the scaling 
variable $x$ and $n_0\,$, $\tau_{z0}$ and $R_0$ are normalization 
constants. We use the convention $n_0 = n(t_0, 0,0,0)$, 
$\tau_{z0}=\tau_z(t_0,r_{z0})$ and $R_0 = R(\tau_{z0})$, where 
$r_{z0}$ is such that, together with $t_0\,$, satisfies 
eq.~(\ref{e:vz}). This implies that $\N(x=0) = 1$. The temperature 
equation, eq. (\ref{e:rT}), is solved by 
\begin{equation} 
T(t,r_x,r_y,r_z) = T_0 
		\left(\frac{\tau_{z0} R_0^2}{\t R^2}\frac{1}{\cosh\xi}\right)^{1/\kappa} 
	\T (x).
\end{equation} 
The constants of normalization are 
$T_0 = T(t_0,0,0,0)$ and $\T (0)= 1$. 
We find that the solution is independent
of the form of the function $\T (x)$, provided that $\T (x) > 0$.

Using the ansatz for the flow profile and the solution for the 
density and the temperature profile, the relativistic Euler 
equation reduces to a complicated non-linear equation that contains 
$R$, $\dot R$ and $\ddot R$ and the variable $x$. Taking this 
equation at $x = 0$ we express $\ddot R$ as a function of $R$ and 
$\dot R$. Substituting this back to the Euler equation we obtain a 
transcendental equation for $\dot R^2$, and $x$.  This equation has 
a particular solution if 
\begin{equation} 
	\dot R = \dot R_0 . \label{rdot}
\end{equation} 
In this case, the acceleration of the radius parameter vanishes, 
$\ddot R = 0$, and the solution is 
$R = R_0 + \dot R_0 (\t - \tau_{z0})$. 
Then the relativistic Euler equation reduces to 
\begin{equation} 
 \left(1 + \frac{1}{\kappa} \right) 
  \left(\frac{R \dot R}{\tau_z} + 3\dot R^2\right) 
	 = 2(1 - x \dot R^2)  \left[\log\N(x) \T(x)\right]^\prime\ , 
 \label{euler}
\end{equation} 
where the lhs depends only on $\t$ while the rhs is only a function
of the variable $x$, hence both sides are constant.
This implies that
\begin{equation} 
	\frac{R}{\t} = \dot R_0\,, \label{e:rsol}
\end{equation} 
thus $R_0 = \dot R_0 \tau_{z0}\,$. Thus the origin of the time axis 
(fixed by the assumption of the scaling longitudinal Bjorken flow 
profile) coincides with the vanishing value of the transverse radius 
parameters. 

The solutions can be casted in a relatively simple form by introducing
the proper time $\tau$,  
\bea
\tau & =&  \sqrt{\t^2 - r_t^2} \, = \, 
	\sqrt{t^2 - r_x^2 - r_y^2 - r_z^2}. \label{e:tau} \\
\eea
Using this natural variable we find that
\bea
	{\mathbf v}  & = & \frac{{\mathbf r}}{t} , \label{e:vsol}\\
	u^\mu & = & \frac{x^\mu}{\tau}.
\eea
Thus the velocity field of our solution corresponds to the flow
field of the spherically symmetric scaling solution. However,
in the scaling solution the temperature and the pressure 
distributions are dependent only on the proper time $\tau$, while 
in our case both the density and the temperature distributions are 
generally dependent on the scale variable $x$ in the  transverse 
direction. 

As the solution is relativistic, and it is defined in the positive light-cone, given by $\tau \ge 0$, we obtain a constraint for the 
transverse coordinate, $r_t \le \t$. This together with 
eq.~(\ref{e:rsol}), the solution for the scale $R$, implies that 
the scaling variable has to satisfy the constraint 
$x \dot R_0^2 \le 1$, which corresponds to the limitation that the 
velocity of the fluid can not exceed the speed of light. 

By replacing eqs.~(\ref{rdot},\ref{e:rsol}) into the Euler equation, eq.(\ref{euler}), one obtains 
\begin{equation} 
\frac{d}{dx}
 \log\left[(1-x\dot R_0^2)^{2(1+1/\kappa)}\N(x)\T(x)\right]=0\,, 
\end{equation} 
which gives, together with the condition $\N(0)\T(0) = 1$,
\begin{equation} 
\N(x) \T(x) = (1 - \dot R_0^2 x)^{-  2(1 + 1/\kappa)}. \label{e:nt}
\end{equation} 
In this family of solutions, the scaling functions for the 
temperature and the density distribution are thus not independent. 
However, a constraint is given for their product, hence one of them
can be chosen as an arbitrary positive function.

For clarity, let us introduce new forms of the scaling functions as 
\bea
	\TT (x) & =& \T (x)  (1 - \dot R_0^2 x)^{2/\kappa}, \\
	\NN (x) & = & \N (x)  (1 - \dot R_0^2 x)^2.
\eea
Then the constraint can be casted to the simple form of
\begin{equation} 
\NN(x) \TT(x) = 1.
\end{equation} 
This construction for the scaling functions of the transverse density
and temperature profiles coincides with the method, that we developed
for the solution of the relativisitic hydrodynamical equations in
the (1+1) dimensional problem, but here the transverse flow 
has a two-dimensional distribution, so the exponents and the scaling
variables had to be re-defined accordingly.

Let us summarize our new family of solutions of the 1+3 dimensional
relativistic hydrodynamics for cylindrically symmetric systems
by substituting the results to the density,
temperature and pressure profiles. 

We obtain
\bea
{\mathbf v} & = & \frac{\mathbf r}{t} , 
		\quad \mbox{\rm for} \quad |{\bf r}| \le t, \\
x & = & \frac{r_t^2}{\dot R^2_0 \t^2}, 
		\quad \mbox{\rm for} \quad {r_t} \le \t, \\
n(t,{\mathbf r}) & = & n_0 \left(\frac{\tau_{z0}}{\tau}\right)^3_0 \NN(x), \label{e:nsol} \\
p(t,{\mathbf r}) & = & p_0 \left(\frac{\tau_{z0}}{\tau}\right)^{3 + 3/\kappa}, \\
T(t, {\mathbf r}) & = & T_0 \left(\frac{\tau_{z0}}{\tau} \right)^{3/\kappa}\frac{1}{\NN (x)}, \label{e:tsol} 
\eea
where $p_0 = n_0 T_0$. 
Note that the scaling variable $x$ is invariant for boosts in the 
longitudinal direction, and it is rotation-invariant in the 
transverse direction, but $x$ is not boost-invariant in the 
transverse directions. Hence we have generated  cylindrically 
symmetric, longitudinally boost invariant solutions of relativistic 
hydrodynamics. In the longitudinal direction, these solutions are 
homogeneous, boost-invariant and also scale-invariant. Due to this 
reason, the observable rapidity distribution is 
\begin{equation} 
	\frac{dn}{dy} = \mbox{\rm const},
\end{equation} 
a flat distribution, corresponding to the ultra-relativistic
nature of the solution in the longitudinal direction (where 
$y = 0.5\log[(E+k_z)/(E-k_z)]$ is the rapidity of a particle with 
four-momentum $(E,{\bf k}) $ and $dn/dy$ is the rapidity 
distribution of particle density).

A new hydrodynamical solution is assigned to each non-negative 
function $\NN (x)$, similarly to the cases of the non-relativistic 
solutions of ref.~\cite{cssol} and the 1+1 dimensional relativistic 
solution of ref.~\cite{s1}. Note that the solutions are 
valid also for massive particles, the form of the solution is 
independent of the value of the mass $m$. The form of solutions 
depends parameterically on $\kappa$, that characterizes the 
equation of state.

We have obtained new solutions of the (1+3) dimensional relativistic 
hydrodynamical equations which describe a self-similar, streaming 
flow. In the case of $\dot R = 1$ and $\NN (x) = 1$ we recover 
the spherically symmetric scale-invariant solution. This means that, 
in this limiting case, the pressure, the density and the temperature 
profiles depend only on the proper time $\tau$. 

In the general case, our solution depends on the scale $R$ (or we 
can say it contains some characteristic scale $R$) and also on an arbitrary scaling function $\NN (x)$.

\section{Summary}
We have found a new family of solutions of 1+3 dimensional 
relativistic hydrodynamics with cyllindrical symmetry. This family 
solves the continuity equation and the conservation of the 
energy - momentum tensor of a perfect fluid, assuming a simple  
equation of state, similarly to the case of the recently found 1+1 
dimensional solutions. The mass of the particles $m$ and the  
constant of proportionality between the kinetic energy density and 
the pressure, $\kappa$, are free parameters of the solution.

As compared to the well-known case of the scale-invariant solution, 
we have solved one more equation, the continuity equation. We have 
considered equations of state that have two free parameters, the 
mass $m$ and $\kappa = \partial \ep/\partial p$, while the 
scale-invariant solution is obtained in the $m=0$ approximation. 
Interestingly, our generalizations resulted in {\it additional 
freedom} in the solution. The new solutions, similarly to the 
scale-invariant case, prescribe scaling 3-dimensional flow and 
pressure distribution. However, in our case, the pressure is a
product of the local number density and the local temperature, 
hence one of these can be chosen in an arbitrary manner.


The essential result of our paper is that we found a rich family of 
exact analytic solutions of relativistic hydrodynamics that 
contain both a longitudinal Bjorken flow (that is frequently utilized
in estimations of observables in high energy heavy ion collisions)
and a relativistic transverse flow (whose existence is evident
from the analysis of the single particle spectra at RHIC and SPS 
energies).

{\it Acknowledgments:}  
T. Cs. would like to thank L.P. Csernai, B. Luk\'acs and 
J. Zim\'anyi for inspiring discussions during the initial phase of 
this work, and to Y. Hama and G. Krein for kind hospitality during 
his stay at USP and IFT, Sao Paulo, Brazil. This work has been 
supported by a NATO Science Fellowship (Cs.T.), by the OTKA grants 
T026435, T029158 and T034269 of Hungary, the NWO - OTKA  grant 
N 25487 of The Netherlands and Hungary, and the grants FAPESP 
00/04422-7, 99/09113-3 of S\~ao Paulo, Brazil. 

\vfill\eject

\vfill\eject


\begin{thebibliography}{99}
\bibitem{Landau} 
	L.D. Landau, Izv. Akad. Nauk SSSR 17 (1953) 51;
	in ``Collected papers of L.D. Landau" (ed. D. Ter-Haar, 
      Pergamon, Oxford, 1965) p. 665 - 700

\bibitem{Bjorken} 
	J.D. Bjorken, Phys. Rev. {\bf D27} (1983) 140

\bibitem{csernai} 
	L.P. Csernai, {\it Introduction to Relativistic Heavy
	Ion Collisions}, John Wiley and Sons, 1994 

\bibitem{biro1}
	T.S. Bir\'o, Phys.Lett. {\bf B474} (2000) 21-26

\bibitem{biro2}
	T.S. Bir\'o, Phys.Lett. {\bf B487} (2000) 133-139

\bibitem{SPH-jpg}
	C.E. Aguiar, T. Kodama, T. Osada, Y. Hama,
	J. Phys. {\bf G27} (2001) 75-94

\bibitem{SPH-qm01}
	C.E. Aguiar, Y. Hama, T. Kodama, T. Osada, Nucl. Phys. 
      {\bf A698} (2002) 639-642. 

\bibitem{jnr}   J. Bondorf, S. Garpman and J. Zim\'{a}nyi, 
                Nucl. Phys. {\bf A296} (1978) 320 .

\bibitem{jde}   J.N. De, S.I.A. Garpman, D. Sperber, J.P. Bondorf 
                and J. Zim\'{a}nyi, Nucl. Phys. {\bf A305} (1978) 
                226.

\bibitem{nr}    T. Cs{\"{o}}rg{\H{o}}, B. L{\"{o}}rstad and 
                J. Zim{\'{a}}nyi; 
                Phys. Lett. {\bf B338} (1994) 134; nucl-th/9408022

\bibitem{nrt}   J. Helgesson, T. Cs\"{o}rg\H{o}, 
                M. Asakawa and B. L\"{o}rstad, 
                Phys. Rev. {\bf C56} (1997) 2626.

\bibitem{cspeter}  
                P. Csizmadia, T. Cs\"{o}rg\H{o} and B. Luk\'{a}cs,
                nucl-th/9805006, Phys. Lett. {\bf B443} (1998) 21.

\bibitem{cssol} T. Cs\"org\H{o}, nucl-th/9809011

\bibitem{ellsol}  S.V. Akkelin, T. Cs\"{o}rg\H{o}, B. Luk\'{a}cs, 
                  Yu.M. Sinyukov and M. Weiner, Phys. Lett. 
                  {\bf B505} (2001), 64.

\bibitem{ellsp}	T. Cs\"{o}rg\H{o}, S.V. Akkelin, Y. Hama,
		      B. Luk\'{a}cs and Yu.M. Sinyukov, 
		      hep-ph/0108067

\bibitem{csell} T. Cs\"org\H{o}, hep-ph/0111139 

\bibitem{3d}    T. Cs\"org\H o and B. L\"orstad,
                Phys. Rev. {\bf C54} (1996) 1390;
	          T. Cs\"org\H o and B. L\"orstad,
                Nucl. Phys. {\bf A590} (1995) 465c.

\bibitem{s1}    T. Cs\"org\H{o}, F. Grassi, Y. Hama and 
                T. Kodama, {\it Simple solutions of 
                relativistic hydrodynamics for 
                longitudinally expanding systems}, 
	          hep-ph/0203204. 

\bibitem{rellsol}	
		T. Cs\"org\H{o}, Y. Hama and T. Kodama,
		{\it Simple solutions of relativistic 
            hydrodynamics for ellipsoidally expanding 
            systems}, manuscript in preparation.

\end{thebibliography}
\end{document}